\documentclass[aps,prl,twocolumn,showpacs,superscriptaddress]{revtex4}
\usepackage[T1]{fontenc}
\usepackage[latin9]{inputenc}
\usepackage{graphicx}
\usepackage{amssymb}



\begin{document}

\title{Circuit QED and sudden phase switching in a superconducting qubit array}

\author{L. Tian}

\email{ltian@ucmerced.edu}

\affiliation{University of California, Merced, 5200 North Lake Road, Merced, CA 95343, USA}

\date{\today}

\begin{abstract}
Superconducting qubits connected in an array can form quantum many-body systems such as the quantum Ising model. By coupling the qubits to a superconducting resonator, the combined system forms a circuit QED system. Here, we study the nonlinear behavior in the many-body state of the qubit array using a semiclassical approach. We show that sudden switchings as well as a bistable regime between the ferromagnetic phase and the paramagnetic phase can be observed in the qubit array. A superconducting circuit to implement this system is presented with realistic parameters . 
\end{abstract}
\maketitle

Circuit quantum electrodynamics (QED) has been intensively explored in the past few years \cite{CQED,CQEDexp}. A number of quantum optical effects such as Lamb shift, nonlinear spectrum, and lasing have been demonstrated in experiments \cite{CQEDrecent}. Most recently, the Tavis-Cummings model has also been tested with multiple transmon qubits coupling to a resonator \cite{TavisCummingsCQED}. The superconducting resonators hence not only provide a powerful tool to implement quantum information protocols \cite{SCircuit,PhotonTransfer,CavityTLS}, but can also facilitate the study of quantum coherence effects and many-body effects involving microwave photons \cite{CCA,CQEDPhaseTransition}. 

When a quantum many-body system is coupled to a resonator cavity, novel many-body phenomena that have not been studied previously in condensed matter systems can be observed. In the atomic systems \cite{CavityQED}, it has been shown that atomic condensate can be coupled strongly to cavity photons to demonstrate interesting effects \cite{AtomCQED, AtomExp}. In the superconducting circuits, given the diversity of the electromagnetic couplings and parameter regimes, a large variety of many-body Hamiltonians can be engineered that can not be emulated in atomic systems. In the past, critical behavior has been studied both theoretically and experimentally in Josephson junction arrays that can emulate spin models and quantum phase models \cite{JJAManyBody}. Recent progresses in superconducting quantum devices further expand such research to include fermionic Hamiltonians \cite{SCircuitQSimu}. 

In this work, we study the nonlinear behavior in the many-body state of an array of superconducting qubits coupled to a superconducting resonator. Different from previous work on optical bistability \cite{QOBook,semiclassical}, our work focuses on the behavior of the qubit array in this system. The qubits are arranged in a one-dimensional chain to form the quantum Ising model. In the bad cavity limit, the photon number of the resonator will depend nonlinearly on the average of the qubit operator. This gives rise to a sudden switching between different quantum many-body phases in the qubit array when the driving power on the resonator is varied, instead of the continuous quantum phase transition in a simple quantum Ising model. Our study also shows that a bistable regime exists where the qubit array can be in either the paramagnetic phase or the ferromagnetic phase. A circuit to implement this model can be constructed with realistic parameters, where the inductive coupling between  the superconducting qubits and the resonator generates a shift of the resonator frequency that depends on the qubit operators. The nonlinear effects studied here can be experimentally demonstrated in a toy system with  only a few qubits and a resonator.  Furthermore, the resonator can act as a threshold detector to measure the phase switching in the bistable regime \cite{ThresholdDetect} and act as a probe of the continuous quantum phase transition in the weak driving limit \cite{CQEDProbe}. 

We consider a qubit array coupled to a resonator with the total Hamiltonian
\begin{equation}
H_{t}=H_{QI}+H_{int}+H_{c0}+H_{\kappa}+H_{\gamma}\label{eq:Ht}
\end{equation}
where $H_{QI}=-J_{x}\sum_{i}\sigma_{xi}-\sum_{i}\sigma_{zi}\sigma_{zi+1}$ describes the quantum Ising model with a uniform transverse magnetic field $J_{x}$ and a ferromagnetic coupling between nearest neighbor qubits, $H_{c0}=\hbar\omega_{c} a^{\dagger}a+\hbar \epsilon(t)(a+a^{\dagger})$ describes the Hamiltonian of the resonator and includes a driving term with the amplitude $\epsilon(t)=2\epsilon\cos\omega t$, and $H_{int}=\hbar ga^{+}a\sum\sigma_{xi}$ is the coupling between the resonator and the qubits. Here, $\sigma_{xi}$ and $\sigma_{zi}$ are the Pauli matrices of the qubits and $a$ ($a^{\dagger}$) is the annihilation (creation) operator for the resonator. The magnitude of the ferromagnetic coupling is set to unity and is used as the energy unit in the following. The term $H_{\kappa}$ ($H_{\gamma}$) is the coupling between the resonator (qubit) and its environmental modes that causes damping (relaxation) of the resonator (qubit). Without the resonator, a second order quantum phase transition between the paramagnetic and the ferromagnetic phases occurs in the qubit array when the transverse field is varied \cite{QPTBook}. 

The coupling $H_{int}$ shifts the frequency of the resonator by $g\sum\sigma_{xi}$. In the Heisenberg picture, we can derive the operator equation for the resonator in the rotating frame, 
\begin{equation}
\dot{a}=i\Delta_{c}a-i\epsilon-iga\sum_{i}(\sigma_{xi})-\frac{\kappa}{2}a+\sqrt{\kappa}a_{in}\label{eq:a_eq}
\end{equation}
where $\Delta_{c}$ is the detuning, $\kappa$ is the damping rate of the resonator, and $a_{in}$ is the noise operator \cite{QOBook}. In the bad cavity limit with strong resonator damping, we can derive the operator average in the steady state as 
\begin{equation}
\langle a\rangle_{ss}=\frac{i\epsilon}{(i\Delta_{c}-\kappa/2)-ig\langle\sum_{i}(\sigma_{xi})\rangle_{ss}},\label{eq:a_ss}
\end{equation}
using a semiclassical approximation with $\langle a\sum_{i}(\sigma_{xi})\rangle_{ss}\approx\langle a\rangle_{ss}\langle\sum_{i}(\sigma_{xi})\rangle_{ss}$, where the quantum correlation between the qubits and the resonator is neglected \cite{semiclassical}. The photon number in the steady state is then
\begin{equation}
n_{s}=\frac{i\epsilon}{\kappa}\langle a-a^{\dagger}\rangle_{ss}=\frac{\epsilon^{2}}{\kappa^{2}/4+(\Delta_{c}-g \langle\sum_{i}(\sigma_{xi})\rangle_{ss})^{2}},\label{nss}\end{equation}
depending nonlinearly on $\langle\sum_{i}(\sigma_{xi})\rangle_{ss}$. By considering the relaxation of the qubit array, we can take the steady state average to be $\langle\sum_{i}(\sigma_{xi})\rangle_{ss}=X(\widetilde{J}_{x})$, where $X$ is the ground state average of $\sum_{i}(\sigma_{xi})$ in the quantum Ising model with the effective transverse field $\widetilde{J}_{x}=J_{x}-\hbar gn_{s}$ and can be calculated with the Jordan-Wigner transformation. Due to the back-action of the resonator, the effective transverse field is now shifted by a term proportional to the photon number. 

\begin{figure}
\includegraphics[width=8.5cm,clip]{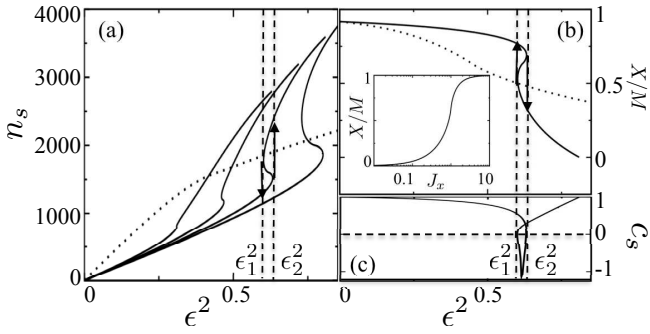} 
\caption{(a) photon number $n_{s}$ vs. driving power $\epsilon^{2}$.  Solid curves from left to right are for $J_{x}=1.4,\,1.6,\,1.8,\,2.0$  at $\Delta_{c}=0$; dotted curve is for $J_{x}=1.8$ at $\Delta_{c}=gM$. (b) $X$ vs. $\epsilon^{2}$ for $J_{x}=1.8$ at $\Delta_{c}=0$ (solid curve) and at $\Delta_{c} = gM$ (dotted curve); inset is $X$ vs. $J_{x}$ in the quantum Ising model. (c) stability coefficient $c_{s}$ vs. $\epsilon^{2}$ for $J_{x}=1.8$ at $\Delta_{c}=0$. We use $\kappa=0.03$, and $g=0.0005$. The dashed lines indicate $\epsilon_{1,2}^{2}$.}
\label{fig1}
\end{figure}
Below we will study how the nonlinear dependence affects the resonator and the quantum many-body system \cite{nonlinearBook}. Choosing $\Delta_{c}=0$ and $J_{x}>1$ to be above the critical point of the quantum Ising model  $J_c=1$, we first plot the photon number $n_{s}$ versus the driving power $\epsilon^{2}$ in Fig.~\ref{fig1}(a). At low driving power, $n_s$ is small so that $\widetilde{J}_{x}\approx J_{x}>1$ and the qubit array is in the paramagnetic phase with $X\approx M$, where M is the number of sites in the array. The photon number increases nearly linearly with $\epsilon^{2}$.  As the driving is increased, $n_s$ becomes larger and $\widetilde{J}_{x}$ decreases to approach the critical point. In this regime, a bistable behavior can be observed: when $\epsilon^{2}$ is increased, $n_s$ initially follows the lower branch but suddenly jumps to the upper branch at $\epsilon_{2}^{2}$; when $\epsilon^{2}$ is decreased from an initial high driving power, $n_s$ follows the upper branch but suddenly jumps to the lower branch at $\epsilon_{1}^{2}$. Meanwhile, the jump in the photon number is accompanied by a sudden switching of the quantum many-body state in the qubit array between the ferromagnetic phase and the paramagnetic phase which can be seen in Fig.~\ref{fig1}(b). Hence, in the bistable regime between $\epsilon_{1}^{2}$ and $\epsilon_{2}^{2}$, the qubit array can be in either the paramagnetic phase or the ferromagnetic phase. Note that the bistability depends critically on the system parameters. For a detuning of $\Delta_{c}=gM$, no bistable regime can be found.

The stability of the semiclassical solutions can be determined from the dynamic equation for the photon number: $dn/dt=i\epsilon(a-a^{\dagger})-\kappa n$ \cite{nonlinearBook}. Assume a small fluctuation $\delta n$ in the photon number $n=n_{s}+\delta n$. The average $X$ obtains a fluctuation $\delta X=-\hbar gX^{\prime}\delta n$ with $X^{\prime}$ being the derivative of $X$. And the fluctuation $\delta(a-a^{\dagger})$ can be derived from Eq. (\ref{eq:a_ss}). We can then derive
\begin{equation}
\frac{d\delta n}{dt}=-\kappa\delta n(1-\frac{2n_{ss}^{2}\hbar g^{3}}{\epsilon^{2}}X^{\prime}X)\label{eq:dn}
\end{equation}
at $\Delta_{c}=0$. When the stability coefficient satisfies $c_{s}=1-2n_{ss}^{2}\hbar g^{3}X^{\prime}X/\epsilon^{2}>0$, the semiclassical solutions are stable \cite{nonlinearBook}. In Fig.~\ref{fig1}(c), $c_{s}$ is plotted which shows that the solutions in both the upper branch and the lower branch are stable while these in the middle branch are unstable with $c_{s}<0$. 

\begin{figure}
\includegraphics[width=8.5cm,clip]{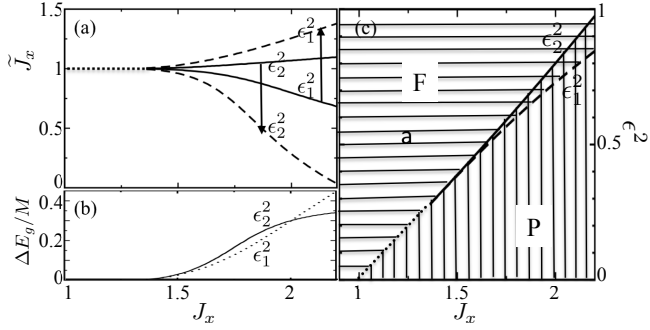} 
\caption{(a) $\widetilde{J}_{x}$ vs. $J_{x}$ before the sudden switching (solid curve) and after the switching (dashed curve) at $\epsilon_{1,2}^{2}$. (b) the energy change at the switching points vs. $J_x$.  (c) many-body phases in the space of $J_{x}$ and $\epsilon^{2}$, where `F' (horizon lines) indicates the ferromagnetic phase and `P' (vertical lines) indicates the paramagnetic phase.}
\label{fig2}
\end{figure}

The effective field $\widetilde{J}_{x}$ of the qubit array also makes a jump from $\widetilde{J}_{x}>1$ to $\widetilde{J}_{x}<1$ at $\epsilon_{2}^{2}$ and vice versa at $\epsilon_{1}^{2}$ as is shown in Fig.~\ref{fig2}(a), when the photon number makes the sudden jump. This shows that the qubit array undergoes a sudden switching between the paramagnetic phase and ferromagnetic phase at $\epsilon_{1,2}^{2}$, and does not cross the point $\widetilde{J}_{x}=1$ continuously. Meanwhile, the ground state energy of the qubit array obtains a finite difference as well. In Fig.~\ref{fig2}(b), we plot the energy difference per qubit $\Delta E_{g}/M$ at the switching points $\epsilon_{1,2}^{2}$ respectively, where the ground state energy of the qubit array at the field $\widetilde{J}_{x}$ is derived using the Jordan-Wigner transformation:
\begin{equation}
E_{g}/M=-(\frac{2}{\pi})(1+\widetilde{J}_{x})E(\frac{4\widetilde{J}_{x}}{(1+\widetilde{J}_{x})^{2}})\label{eq:eg}
\end{equation}
with $E(x)$ being the complete elliptic integral of the second kind. Such behavior in this driven system is in analogy to the first order phase transition in an equilibrium system. While in a simple quantum Ising model without the resonator, continuous quantum phase transition occurs at the critical point $J_{c}=1$. 

The boundaries of the phase switchings $\epsilon_{1,2}^{2}$ as a function of $J_{x}$ can be derive by seeking local maxima and local minima in the curve of $\epsilon^{2}$ versus $n_{s}$ for each $J_{x}$. Using the phase boundaries, we can characterize the quantum many-body phases of the qubit array in terms of two parameters: the transverse field $J_{x}$ and the driving power $\epsilon^{2}$, as is shown in Fig.\ref{fig2}(c). For $\epsilon^{2}>\epsilon_{2}^{2}(J_{x})$, the system is always in the ferromagnetic phase; for $\epsilon^{2} < \epsilon_{1}^{2}(J_{x})$, the system is always in the paramagnetic phase. In the overlapping area between these two boundaries, both phases can exist depending on the history of the parameters. The coupled system of the qubit array and the resonator can thus be explored to demonstrate nonlinear effects in the quantum many-body state. 

\begin{figure}
\includegraphics[width=8.5cm,clip]{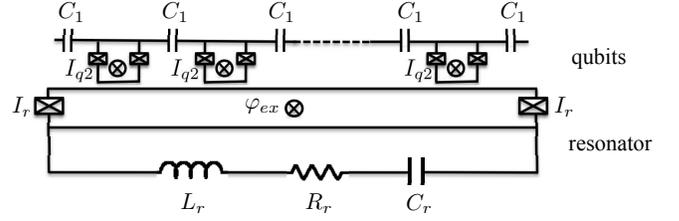}
\caption{Circuit of an array of superconducting qubits coupling to a resonator. The current in the qubit loops couples to the SQUID in the resonator via mutual inductance.}
\label{fig3}
\end{figure}

The quantum Ising model can be constructed with superconducting qubits \cite{SCircuit} connected in a chain and coupled capacitively to their neighboring qubits, as is shown in Fig.~\ref{fig3}. The Hamiltonian can be written as
\begin{equation}
H_{QI}=2e^{2}\vec{n}^{T}C^{-1}\vec{n}-\sum E_{Ji}\cos(\varphi_{i})\label{eq:H0}
\end{equation}
where $\vec{n}^{T}=[n_{1}-(C_{g1}V_{g1}/2e),\cdots,n_{i}-(C_{gi}V_{gi}/2e),\cdots]$ is the number of Cooper pairs on the superconducting islands biased by the gate charge number $C_{gi}V_{gi}/2e$, and $C^{-1}$ is the inverse of the capacitance matrix with $C_{ii}=C_{0}$ being the total capacitance connected to the $i$-th island and $C_{i\, i\pm1}=-C_{1}$ being the coupling capacitance between two adjacent islands with $C_{1}<C_{0}$. Expanded to the second order of $C_{1}/C_{0}$, the elements of $C^{-1}$ can be derived as: $C_{ii}^{-1}=1/C_{0}+2C_{1}^{2}/C_{0}^{3}$, $C_{i\, i+1}^{-1}=C_{1}/C_{0}^{2}$, and $C_{i\, i+n}^{-1}=o(C_{1}/C_{0}^{2})(1/C_{0})$ for $n>1$. To obtain the ferromagnetic coupling, we define the spin states of the $i$-th qubit by the charge states $|0\rangle_{i}$ and $|1\rangle_{i}$ on the $i$-th island as: $|0\rangle_{i}\rightarrow|\uparrow\rangle_{i}, |1\rangle_{i}\rightarrow|\downarrow\rangle_{i}$ when $i$ is an odd number; and $|1\rangle_{i}\rightarrow|\uparrow\rangle_{i}, |0\rangle_{i}\rightarrow|\downarrow\rangle_{i}$ when $i$ is an even number. By setting the gate voltages to be at the degeneracy point with $C_{gi}V_{gi}/2e=1/2$, the single qubit term in Eq. (\ref{eq:H0}) becomes $\sum(-E_{Ji}/2)\sigma_{xi}$; and the coupling term can be written as $\sum(-B_{1i}\sigma_{zi}\sigma_{zi+1}+B_{2i}\sigma_{zi}\sigma_{zi+2})$ to the second order of $C_{1}/C_{0}$ with $B_{1i}=4e^{2}(C_{1}/C_{0})$ and $B_{2i}=4e^{2}(C_{1}/C_{0})^{2}$. Due to the long range nature of the Coulomb interaction, the coupling term includes not only the nearest neighbor couplings, but also unwanted couplings between other qubits. However, it has been shown that the phase diagram of this model is qualitatively the same as that of the simple quantum Ising model when $B_{2i}/B_{1i}=C_{1}/C_{0}< 1/2$ \cite{NNN}. In our system, $C_{1}/C_{0}$ can be readily designed to satisfy this criterion.

To generate a coupling $H_{int}=\hbar ga^{+}a\sum\sigma_{xi}$ between the qubits and the resonator, we consider a resonator that includes a SQUID in the circuit. In Fig.~\ref{fig3}, each qubit has a small loop current $I_{qi}=I_{q2}\sin(\delta_{2}/2)\sigma_{xi}$ \cite{Quantronium} that is controlled by the external flux $\delta_{2}$. This current is coupled to the SQUID via mutual inductance and modifies the flux in the SQUID loop by $\Phi_{r}=-\mu_{0}I_{q2}R_{0} \sin(\delta_{2}/2)\sum_{i}\sigma_{xi}$ where $R_{0}$ is roughly the size of the qubit loop. The effective inductance of the SQUID can then be written as 
\begin{equation}
L_{sq}=(\frac{\hbar}{2e})\frac{1}{2I_{r}\cos\varphi_{ex}-2I_{r}\sin\varphi_{ex}(\pi\Phi_{r}/\Phi_{0})}\label{eq:Lsq}
\end{equation}
where $I_{r}$ is the critical current of the SQUID junctions and $\varphi_{ex}$ is the external flux in the SQUID loop. The total inductance of the resonator is $L_{tot}=L_{sq}+L_{r}$, including the SQUID inductance and the loop inductance $L_{r}$. Let $L_{sq}^{0}$ be the SQUID inductance at $\Phi_{r}=0$ and $\varphi_{ex}=\pi/4$, the frequency of the resonator becomes
\begin{equation}
\omega_{c}=\omega_{c0}+\omega_{c0}(\frac{L_{sq}^{0}}{L_{sq}^{0}+L_{r}})(\frac{\pi\mu_{0}R_{0}I_{q2}}{2\sqrt{2}\Phi_{0}})\sum_{i}\sigma_{xi}\label{eq:wc}
\end{equation}
with $\omega_{c0}=((L_{sq}^{0}+L_{r})C_{r})^{-1/2}$. This expression includes the coupling between the qubits and the resonator in the form of $H_{int}$ with 
\begin{equation}
g=\omega_{c0}(\frac{L_{sq}^{0}}{L_{sq}^{0}+L_{r}})(\frac{\pi}{2})(\frac{\mu_{0}R_{0}I_{q2}}{\Phi_{0}\sqrt{2}}).\label{eq:g}
\end{equation}
The coupling constant $g$ depends on the qubit flux $\Phi_{r}$ to the first order, which is a merit of this circuit that ensures a sizable coupling to demonstrate the nonlinear effect. We can choose the following parameters: $L_{r}\sim100\,\textrm{pH}$, $I_{q2}=80\,\textrm{nA}$, $I_{r}=1200\,\textrm{nA}$ (i.e. $L_{sq}^{0}=100\,\textrm{pH}$), and the total capacitance of the resonator circuit $C_{r}=0.1\,\textrm{pF}$, which give $g=2\pi\times1\,\textrm{MHz}$ and $\omega_{c}=2\pi\times29\,\textrm{GHz}$. With a ferromagnetic coupling of $2\pi\times2\,\textrm{GHz}$ chosen as the energy unit, we have $g=0.0005$ in the relative unit. Note that various other qubits such as the superconducting flux qubit and phase qubit can be used to construct the qubit array given the flexibility in the superconducting circuits \cite{QIsingLevitov,FluxQ,PhaseQ}.

Decoherence plays an essential role in observing the nonlinear effect. In the above, we assume that the resonator is in the bad cavity limit where the damping of the resonator is stronger than the coupling constant and the decoherence of the qubits. This can be achieved either by using a noisy resonator or by connecting the resonator to a dissipative element which is shown as the resistor $R_{r}$ in Fig.~\ref{fig3}. The decoherence of the qubits will relax the quantum many-body system to its ground state at the effective transverse field. Given the sub-megahertz decoherence rates that are recently measured in the superconducting qubits, the nonlinear effect can be studied in the above circuit with realistic parameters. Finite temperature may play a role in this system as well when the field $\widetilde{J}_{x}$ becomes much smaller than the ferromagnetic coupling, where a low-lying excited state is nearly degenerate with the ground state. The final state can then be a mixture of the ground state and the excited state at a temperature of $T=20\,\textrm{mK}$ ($k_{B}T/\hbar=2\pi\times0.4\,\textrm{GHz}$). In addition, the detection of the phase switchings in the qubit array can be achieved by measuring the photon number in the resonator. The jump of the photon number is directly associated with the phase switching when the driving is in the bistable regime. 

To conclude, we studied the nonlinear effect in the quantum many-body state of a superconducting qubit array coupling to a superconducting resonator. We showed that sudden phase switching between the paramagnetic phase and the ferromagnetic phase can occur in the qubit array when the driving on the resonator is varied due to the strong back-action of the resonator on the qubits. The phase switching results in a finite energy jump in the qubit array, in analogy to a first order phase transition. These results showed that novel phenomena can be observed when a global resonator cavity is connected to a quantum many-body system. Many interesting effects await to be discovered in such systems. 

\emph{Acknowledgements.} The author would like to thank Prof. R. Chiao for valuable discussions. This work is supported by the National Science Foundation under Grant No. NSF-CCF-0916303 and by UC Merced funds.

\end{document}